\documentclass[11pt,twoside]{article}
\usepackage{asp2010}

\resetcounters

\begin{document}

\title{The MUCHFUSS project - Searching for the most massive companions to hot subdwarf stars in close binaries and finding the least massive ones}
\author{S. Geier$^1$, V. Schaffenroth$^1$, H. Hirsch$^1$, A. Tillich$^1$, U. Heber$^1$, L. Classen$^1$, T. Kupfer$^{1,2}$, P. F. L. Maxted$^3$, R. H. \O stensen$^4$, B. N. Barlow$^5$, S. J. O'Toole$^6$, T. R. Marsh$^7$, B. T. G\"ansicke$^7$, O. Cordes$^8$, R. Napiwotzki$^9$
\affil{$^1$Dr. Karl Remeis-Observatory \& ECAP, Astronomical Institute,
Friedrich-Alexander University Erlangen-Nuremberg, Sternwartstr. 7, D 96049 Bamberg, Germany}
\affil{$^2$Department of Astrophysics, Faculty of Science, Radboud University Nijmegen, P.O. Box 9010, 6500 GL Nijmegen, NE}
\affil{$^3$Astrophysics Group, Keele University, Staffordshire, ST5 5BG, UK}
\affil{$^4$Institute of Astronomy, K.U.Leuven, Celestijnenlaan 200D, B-3001 Heverlee, Belgium}
\affil{$^5$Department of Physics and Astronomy, University of North Carolina, Chapel Hill, NC 27599-3255, USA}
\affil{$^6$Australian Astronomical Observatory, PO Box 296, Epping, NSW, 1710, Australia}
\affil{$^7$Department of Physics, University of Warwick, Conventry CV4 7AL, UK}
\affil{$^8$Argelander-Institut f\"ur Astronomie, Auf dem H\"ugel 71, 53121 Bonn, Germany}
\affil{$^9$Centre of Astrophysics Research, University of Hertfordshire, College Lane, Hatfield AL10 9AB, UK}}

\begin{abstract}
The project Massive Unseen Companions to Hot Faint Underluminous Stars from SDSS (MUCHFUSS) aims at finding hot subdwarf stars with massive compact companions (massive white dwarfs $M>1.0\,{\rm M_{\odot}}$, neutron stars or stellar mass black holes). The existence of such systems is predicted by binary evolution theory and some candidate systems have been found. We classified about $\simeq1400$ hot subdwarf stars from the Sloan Digital Sky Survey (SDSS) by colour selection and visual inspection of their spectra. Stars with high velocities have been reobserved and individual SDSS spectra have been analysed. In total $201$ radial velocity variable subdwarfs have been dis\-covered and about $140$ of them have been selected as good candidates for follow-up time resolved spectroscopy to derive their orbital parameters and photometric follow-up to search for features like eclipses in the light curves. Up to now we found seven close binary sdBs with short orbital periods ranging from $\simeq0.21\,{\rm d}$ to $1.5\,{\rm d}$ and two eclipsing binaries with companions that are most likely of substellar nature. A new pulsating sdB in a close binary system has been discovered as well.
\end{abstract}

\section{Introduction}

A large fraction of the sdB stars ($\simeq50\%$) are short period binaries \citep{maxted01,napiwotzki04} with periods ranging from only $0.07\,{\rm d}$ to more than $10\,{\rm d}$. Close binary sdBs are most likely formed by common envelope (CE) ejection \citep{han02,han03}. However, it is difficult to determine the nature of the close companions in sdB binaries. Because most of them are single-lined, only lower mass limits have been derived from the binary mass functions, which are in general compatible with late main sequence stars of spectral type M or compact objects like white dwarfs. Only in rare cases (e.g. eclipsing systems) it is possible to distinguish between these two options. 

Subdwarf binaries with massive WD companions turned out to be candidates for supernova type Ia (SN Ia) progenitors because these systems lose angular momentum due to the emission of gravitational waves and shrink. Mass transfer or the subsequent merger of the system may cause the WD to reach the Chandrasekhar limit and explode as a SN~Ia. One of the best known candidate systems for the double degenerate merger scenario is the sdB+WD binary KPD\,1930$+$2752 \citep{maxted00,geier07}. 

\citet{geier10a,geier10b} analysed high resolution spectra of single-lined sdB binaries. Because the inclinations of these systems are unknown, additional information is required to derive masses. \citet{geier10a,geier10b} measured the surface gravities and projected rotational velocities. Assuming synchronised orbits the masses and the nature of the unseen companions was constrained. Surprisingly, some companions may be either massive white dwarfs, neutron stars (NS) or stellar mass black holes (BH). However, the assumption of orbital synchronisation in close sdB binaries was shown to be not always justified and the analysis suffers from selection effects \citep{geier10b}. The existence of sdB+NS/BH systems is predicted by binary evolution theory \citep{podsi02,pfahl03,yungelson05,nelemans10}. The formation channel includes two phases of unstable mass transfer and one supernova explosion and the fraction of those systems is consistently predicted to be about $1-2\%$. If the companion were a neutron star, it could be detectable by radio observations as a pulsar. \citet{coenen11} searched for pulsed radio emission at the positions of four candidate systems from \citet{geier10b} using the Green Bank radio telescope, but did not detect any signals. 

We started a radial velocity (RV) survey (Massive Unseen Companions to Hot Faint Underluminous Stars from SDSS\footnote{Sloan Digital Sky Survey}, MUCHFUSS) to find sdBs with compact companions like supermassive white dwarfs ($M>1.0\,M_{\rm \odot}$), neutron stars or black holes \citep{geier11a,geier11b}. The same selection criteria that we applied to find such binaries are also well suited to single out hot subdwarf stars with constant high radial velocities in the Galactic halo like extreme population II and hypervelocity stars \citep[see Heber et al. these proceedings; ][]{tillich11}.

\section{Extended target selection}

\subsection{Colour and RV selection}

For the MUCHFUSS project the target selection is optimised to find massive compact companions in close orbits around sdB stars \citep[for details see][]{geier11a}. The SDSS spectroscopic database is the starting point for our survey. While the target selection presented in \citet{geier11a} includes SDSS Data Release 6, we have now extended the selection to Data Release 7. Hot subdwarf candidates were selected by applying a colour cut to SDSS photometry. All point source spectra within the colours $u-g<0.4$ and $g-r<0.1$ were selected and downloaded from the SDSS Data Archive Server\footnote{das.sdss.org}. By visual inspection we selected and classified $\simeq10\,000$ hot stars. The sample contains $1369$ hot subdwarfs. 

We excluded sdBs with radial velocities (RVs) lower than $\pm100\,{\rm km\,s^{-1}}$ to filter out such binaries with normal disc kinematics, by far the majority of the sample. Another selection criterion is the brightness of the stars. Most objects much fainter than $g=19\,{\rm mag}$ have been excluded. 

\subsection{Survey for RV variable stars}

Second epoch medium resolution spectroscopy ($R=1800-4000$) was obtained using ESO-VLT/FORS1, WHT/ISIS, CAHA-3.5m/TWIN and ESO-NTT/EFOSC2. Up to now we have reobserved $88$ stars. Second epoch observations by SDSS have been used as well. We discovered $58$ RV variable systems in this way. 

\begin{figure}
\begin{center}
  \includegraphics[width=10cm]{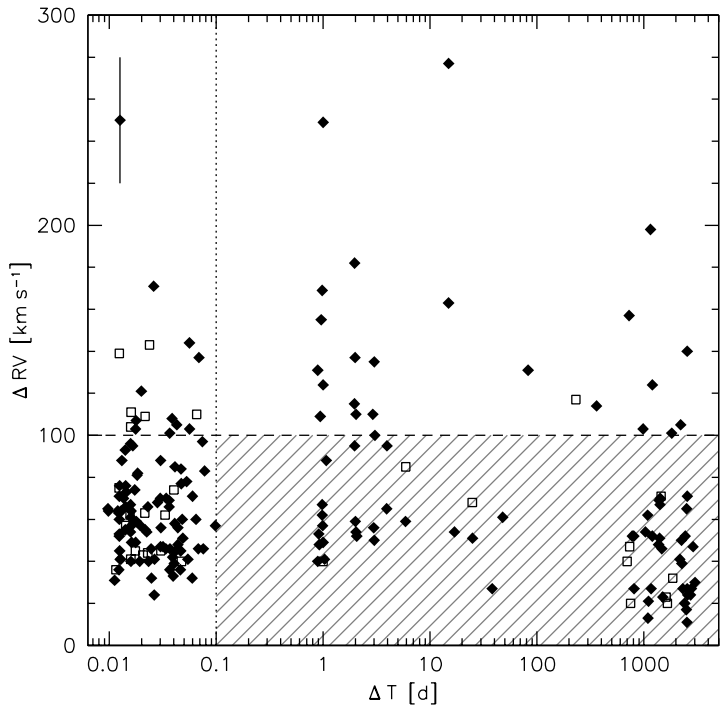}
\end{center}
\caption{Highest radial velocity shift between individual spectra ($\Delta RV$) plotted against time difference between the corresponding observing epochs ($\Delta T$). The dashed horizontal line marks the selection criterion $\Delta RV>100\,{\rm km\,s^{-1}}$, the dotted vertical line the selection criterion $\Delta T<0.1\,{\rm d}$. All objects fulfilling at least one of these criteria lie outside the shaded area and belong to the top candidate list for the follow-up campaign. The filled diamonds mark sdBs, while the open squares mark He-sdOs.}
\label{fig:dTdRV}
\end{figure}

The SDSS spectra are co-added from at least three individual ``sub-spectra'' with typical exposure times of $15\,{\rm min}$ taken consecutively. Hence, SDSS spectroscopy can be used to probe for radial velocity variations on short timescales. We have obtained the sub-spectra for all sdBs brighter than $g=18.5\,{\rm mag}$. From the inspection of these data, we discovered $143$ new sdB binaries with radial velocity variations on short time scales ($\simeq0.03\,{\rm d}$). In total we found $201$ new RV variable hot subdwarf stars (see Fig.~\ref{fig:dTdRV}).

In addition $30$ He-sdOs show signs of RV variability. This fraction was unexpected since in the SPY sample only $4\%$ of these stars turned out to be RV variable \citep{napiwotzki08}. However, it is not yet clear what causes this RV variability. Up to now we were not able to derive the orbital parameters of any such object and prove, that it is a close binary star. 

\subsection{Selection of candidates with massive companions}

In order to select the most promising targets for follow-up, we carried out numerical simulations and estimated the probability for a subdwarf binary with known RV shift to host a massive compact companion. We created a mock sample of sdBs with a close binary fraction of $50\,\%$ and adopted the distribution of orbital periods of the known sdB binaries. Two RVs were taken from the model RV curves at random times and the RV difference was calculated for each of the $10^{6}$ binaries in the simulation sample. Since the individual SDSS spectra were taken within short timespans, another simulation was carried out, where the first RV was taken at a random time, but the second one just $0.03\,{\rm d}$ later. Our simulation gives a quantitative estimate based on our current knowledge of the sdB binary populations \citep[for details see][]{geier11a}. 

The extended sample of promising targets including SDSS DR7 consists of $140$ objects in total. These objects either show significant RV shifts ($>30\,{\rm km\,s^{-1}}$) within $0.03\,{\rm d}$ ($114$ stars) or high RV shifts ($100-300\,{\rm km\,s^{-1}}$) within more than one day ($26$ stars). 

\section{Sample statistics}

The classification of the hot subdwarf sample is based on existence, width, and depth of helium and hydrogen absorption lines as well as the flux distribution between $4000$ and $6000\,{\rm \AA}$. Subdwarf B stars show broadened hydrogen Balmer and He\,{\sc i} lines, sdOB stars He\,{\sc ii} lines in addition, while the spectra of sdO stars are dominated by weak Balmer and strong He\,{\sc ii} lines depending on the He abundance. A flux excess in the red compared to the reference spectrum as well as the presence of spectral features such as the Mg\,{\sc i} triplet at $5170\,{\rm \AA}$ or the Ca\,{\sc ii} triplet at $8650\,{\rm \AA}$ were taken as indications of a late type companion.

In total we found $1369$ hot subdwarfs, consistent with the preliminary number of hot subdwarfs ($1409$) found by \citet{kleinman10} in SDSS-DR7. $983$ belong to the class of single-lined sdBs and sdOBs.  Features indicative of a cool companion were found for $98$ of the sdBs and sdOBs. $9$ sdOs have main sequence companions, while $262$ sdOs, most of which show helium enrichment, are single-lined.

The fraction of close binaries among the hot subdwarf stars in SDSS can be estimated by taking a look at the objects with more than one epoch of spectroscopy. $52$ stars ($34$ sdB/sdOB, $7$ He-sdO, $11$ sdB+MS) from our sample have at least two epochs of observations. $53\%$ of the sdBs and sdOBs are RV variable, while only one He-sdO ($\simeq14\%$) and one sdB with a  visible companion ($\simeq9\%$) show variability. Due to the small sample size the last two numbers should be regarded as upper limits at most. The binary fraction of the sdB stars is closer to the one found in the SPY project \citep[$\simeq40\%$][]{napiwotzki04} than to the higher fraction of $\simeq70\%$ reported by \citet{maxted01}. 

\section{Spectroscopy follow-up}

Follow-up medium resolution ($R=1200-4000$) spectra were taken during dedicated follow-up runs with ESO-NTT/EFOSC2, WHT/ISIS, CAHA-3.5m/TWIN, INT/IDS, SOAR/Goodman and Gemini-N/GMOS. Orbital parameters of eight sdB binaries discovered in the course of the MUCHFUSS project have been determined so far \citep{geier11b,geier11c}. 

Since the programme stars are single-lined spectroscopic binaries, only their mass functions $f_{\rm m} = M_{\rm comp}^3 \sin^3i/(M_{\rm comp} + M_{\rm sdB})^2 = P K^3/2 \pi G$ can be calculated. Although the RV semi-amplitude $K$ and the period $P$ can be derived from the RV curve, the sdB mass $M_{\rm sdB}$, the companion mass $M_{\rm comp}$ and the inclination angle $i$ remain free parameters. Adopting the canonical mass for core helium-burning stars $M_{\rm sdB}=0.47\,M_{{\rm \odot}}$ and $i<90^{\rm \circ}$ we derive a lower limit for the companion mass. 

Depending on this minimum mass a qualitative classification of the companions' nature is possible in certain cases. For mini\-mum companion masses lower than $0.45\,M_{\rm \odot}$ a main sequence companion can not be excluded because its luminosity would be too low to be detectable in the optical spectra \citep{lisker05}. In this case the companion can be a compact object like a WD or a late main sequence star. If the minimum companion mass exceeds $0.45\,M_{\rm \odot}$ and no spectral signatures of the companion are  visible, it must be a compact object. If this mass limit exceeds $1.00\,M_{\rm \odot}$ or even the Chandrasekhar limit ($1.40\,M_{\rm \odot}$) the existence of a supermassive WD or even an NS or BH companion is proven.

The minimum companion masses of seven binaries are similar ($0.32-0.41\,M_{\rm \odot}$). From these minimum masses alone the nature of the companions cannot be constrained unambiguously. However, the fact that all seven objects belong to the sdB binary population with the highest minimum masses illustrates that our target selection is efficient and singles out sdB binaries with massive companions \citep{geier11b}. 

\section{Photometry follow-up}

Photometric follow-up allows us to clarify the nature of the companions. Short period sdB binaries with late main sequence or substellar companions show variability in their light curves caused by the irradiated surfaces of the cool companions facing the hot subdwarf stars. If this so-called reflection effect is present, the companion is most likely a main sequence star. If not, the companion is most likely a compact object. In the case of the short period system J1138$-$0035 a light curve taken by the SuperWASP project shows no variation exceeding $\simeq1\%$. The companion is therefore most likely a white dwarf \citep{geier11b}.

We obtained follow-up photometry with the Mercator telescope and the BUSCA instrument mounted on the CAHA-2.2m telescope. In this way we discovered the first eclipsing sdB binary J082053.53+000843.4 to host a brown dwarf companion with a mass ranging from $0.045$ to $0.068\,M_{\rm \odot}$ \citep{geier11c}. 

\begin{figure}
\begin{center}
  \includegraphics[width=8cm, angle=-90]{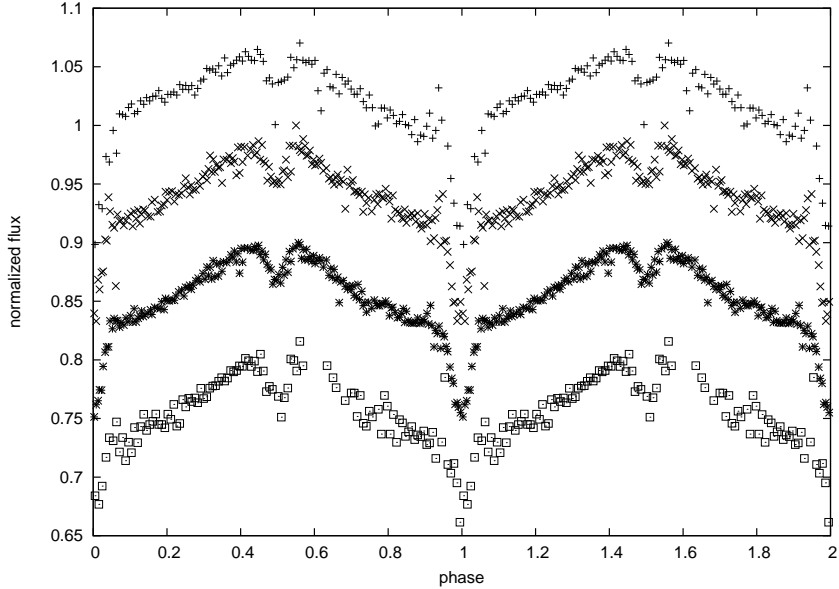}
\end{center}
\caption{Phased light curves of J162256.66+473051.1 taken with BUSCA (UV,B,R,IR-band). Primary and secondary eclipses can be clearly seen as well as the sinusoidal shape caused by the reflection effect.}
\label{fig:J1622}
\end{figure}

The very similar eclipsing system J162256.66+473051.1 was discovered serendipituously (see Fig.~\ref{fig:J1622}). A preliminary analysis shows that the orbital period is very short ($\simeq0.07\,{\rm d}$) and the RV semi-amplitude quite low ($\simeq47\,{\rm km\,s^{-1}}$). The companion is most likely a substellar object as well. The high success rate in finding these objects shows that our target selection not only singles out sdB binaries with high RV-amplitudes, but also systems with very short orbital periods. Low-mass stellar and substellar companions may yet play an underestimated role in the formation of sdB stars (see Geier et al. these proceedings).

Most recently, we detected p-mode pulsations in the sdB J012022.94+395059.4 \citep[FBS\,0117+396, ][]{geier11a} as well as a longer trend indicative of a reflection effect in a light curve taken with BUSCA. Only a few of the known short-period sdB pulsators (sdBV$_{\rm r}$) are in close binary systems. More observations are needed to determine the orbital parameters of this system.

\section{Summary}

The MUCHFUSS project aims at finding hot subdwarf stars with massive compact companions. We identified  1369 hot subdwarfs by colour selection and visual inspection of the SDSS-DR7 spectra. The best candidates for massive compact companions are followed up with time resolved medium resolution spectroscopy. Up to now orbital solutions have been found for eight single-lined binaries. Seven of them have large minimum companion masses compared to the sample of known close binaries, which shows that our target selection works quite well. However, it turns out that our selection strategy also allows us to detect low-mass companions to sdBs in very close orbits. We discovered an eclipsing sdB with a  brown dwarf companion and a very similar candidate system in the course of our photometric follow-up campaign. These early results encourage us to go on, because they demonstrate that MUCHFUSS will find both massive and substellar companions to sdB stars.


\end{document}